\documentclass[aps,pra,nofootinbib,preprint,amsmath,amssymb,floatfix]{revtex4}
\usepackage{color}
\usepackage{graphicx}

\begin{document}

\newcommand{\tc}{\textcolor}
\newcommand{\g}{blue}
\newcommand{\ve}{\varepsilon}
\title{A critical discussion of different methods and models in Casimir effect}
\author{ Iver Brevik$^1$  }      
\affiliation{$^1$Department of Energy and Process Engineering, Norwegian University of Science and Technology, N-7491 Trondheim, Norway}
\date{\today}          
\author{Boris Shapiro$^2$ }
\affiliation{$^2$  Department of Physics, Technion, Israel Institute of Technology, Haifa, Israel }

\begin{abstract}

The Casimir-Lifhitz force acts between neutral material bodies and is due to the fluctuations (around zero) of the electrical polarizations of the bodies. This force is a macroscopic manifestation of the van der Waals forces between atoms and molecules. In addition to being of fundamental interest, the Casimir-Lifshitz force plays an important role in surface physics, nanotechnology and biophysics.
There are two different approaches in the theory of this force. One is centered on the fluctuations inside the bodies, as the source of the fluctuational electromagnetic fields and forces. The second approach is based on finding the eigenmodes of the field, while the material bodies are assumed to be passive and non-fluctuating. In spite of the fact that both approaches have a long history, there are still some misconceptions in the literature. In particular, there are claims that (hypothetical) materials with a strictly real dielectric function $\varepsilon(\omega)$ can give rise to fluctuational Casimir-Lifshitz forces. We review and compare the two approaches, using the simple example of the force in the absence of retardation. We point out that also in the second (the "field-oriented") approach one cannot avoid introducing an infinitesimal imaginary part into the dielectric function, i.e. introducing some dissipation. Furthermore, we emphasize that the requirement of analyticity of  $ \varepsilon(\omega)$ in the upper half of the complex $\omega$ plane is not the only one for a viable dielectric function. There are other requirements as well. In particular, models that use a strictly real  $\varepsilon(\omega)$ (for all real positive $\omega)$ are inadmissible and lead to various contradictions and inconsistencies. Specifically, we present a critical discussion of the "dissipation-less plasma model". Our  emphasis is not on the most recent developments in the field but on some conceptual, not fully resolved issues.
\end{abstract}
\maketitle

\bigskip
\section{Introduction}
\label{secintro}

\text It is well known that material bodies in thermal equilibrium with the environment at some temperature $T$ exert long-range attractive forces on each other. The bodies are electrically neutral and do not possess a permanent dipole (or any higher multipole) moment, so the forces are due solely to the fluctuating electromagnetic fields which are always present (thermal equilibrium conditions assumed). Such forces are often called van der Waals forces. Interchangeably, they are also known as  Casimir-Lifshitz  forces.
Some general treatises on the Casimir effect from various perspectives, can be found in Refs.~\cite{buhmann12,bordag09,milton01,milton04,sernelius18,ellingsen08,plunien86,brevik14,barash89,barash75,landau80,lifshitz61,landau60,rytov89,bimonte21,vankampen68,li19,ginzburg79,col,belin,lam,lam',EJP,steve,PTbrevik,steve1,stan}. In the present analysis we will not consider the Casimir effect in general, but focus on the following  issue.

 There are essentially two different ways to proceed when encountering  the Casimir effect. The first one has its root in quantum statistical mechanics \cite{landau80,lifshitz61}, and consists in  regarding the force to arise  from the fluctuations of the dipoles in the media. The fluctuation-dissipation theorem (FDT) plays here a central role and the dielectric function $\epsilon(\omega)$ must, of course, contain an imaginary part.   The second approach, falling into line   with the original Casimir work \cite{casimir48}, is to consider the problem as  a quantum field theoretical problem, implying that one starts from the electromagnetic field eigenmodes in the system. The total field energy is then obtained by summing over all eigenvalues (to have real eigenvalues one must assume a strictly real $\epsilon(\omega)$ ).
 We will refer to this approach as the quantum field theory (QFT) approach.
 It  was introduced in connection with the Casimir effect in Ref.~\cite{vankampen68} and
 it has proved to be a valuable method in a variety of cases.
There is yet another, "scattering approach" \cite{lam,lam'} to the problem. In the latter, unlike the QFT approach, the system is made open and instead of the "cavity eigenmodes" of QFT one introduces scattering states. We do not consider this useful approach in the present paper.


The derivations of the Casimir-Lifshitz force presented in textbooks on theoretical physics \cite{landau80,landau60} or the reviews \cite{barash89,lifshitz61,rytov89} usually deal with the most general case and do not attempt to compare between the FDT and QFT approaches.
One of the aims of the present paper is to present a simple derivation of the Casimir-Lifhitz force, using the FDT approach in the non-retarded limit, and to compare the result with that obtained within the QFT approach (Sec. II-IV).
We emphasize that, although the eigenmodes used in the QFT approach are well defined only if the medium is dissipation-less, one still must introduce an infinitesimal dissipation $(Im\epsilon(\omega)>0)$ when calculating the Casimir-Lifshitz force. This fact is not always appreciated in the literature. Furthermore, in Sec.IV we discuss some models of material bodies, employed in the theory of Casimir-Lifshitz forces, and point out that models with strictly real $\epsilon(\omega)$ are inadmissible idealizations. Such models violate some basic physical requirements that any material must satisfy. In Sec. V we  elaborate on the limit when both the frequency $\omega$ and the dissipation $(Im\epsilon(\omega))$ become very small. There have been claims that in this limit the  theory, based on the standard  Drude model, breaks down and is in conflict with the Nernst heat theorem. We argue that in this limit the local dielectric function  $\epsilon(\omega)$ loses its meaning and the (non-local) spatial dispersion effects become unavoidable. In Sec. VI the  Drude model with spatial dispersion is briefly discussed and our conclusions are summarized in Sec. VII.

   \section{The FDT approach}

The quantum statistical mechanical approach  is a general and rigorous approach that has also a great intuitive appeal. All material bodies possess fluctuating charges and currents whose spectral density (in equilibrium) is determined by the FDT. It is these currents, and the corresponding fluctuating electromagnetic fields, that give rise to  the Casimir-Lifshitz  forces. The FDT approach, originally due to Rytov, is often called "fluctuational electrodynamics" (see \cite{landau60} for a nice presentation and \cite{rytov89} for a later review).  In this section we outline the approach, using the standard  setup of two dielectric half-spaces separated by a gap of width $l$ (Fig. 1).

\begin{figure}[h]
\includegraphics[width=4.5in]{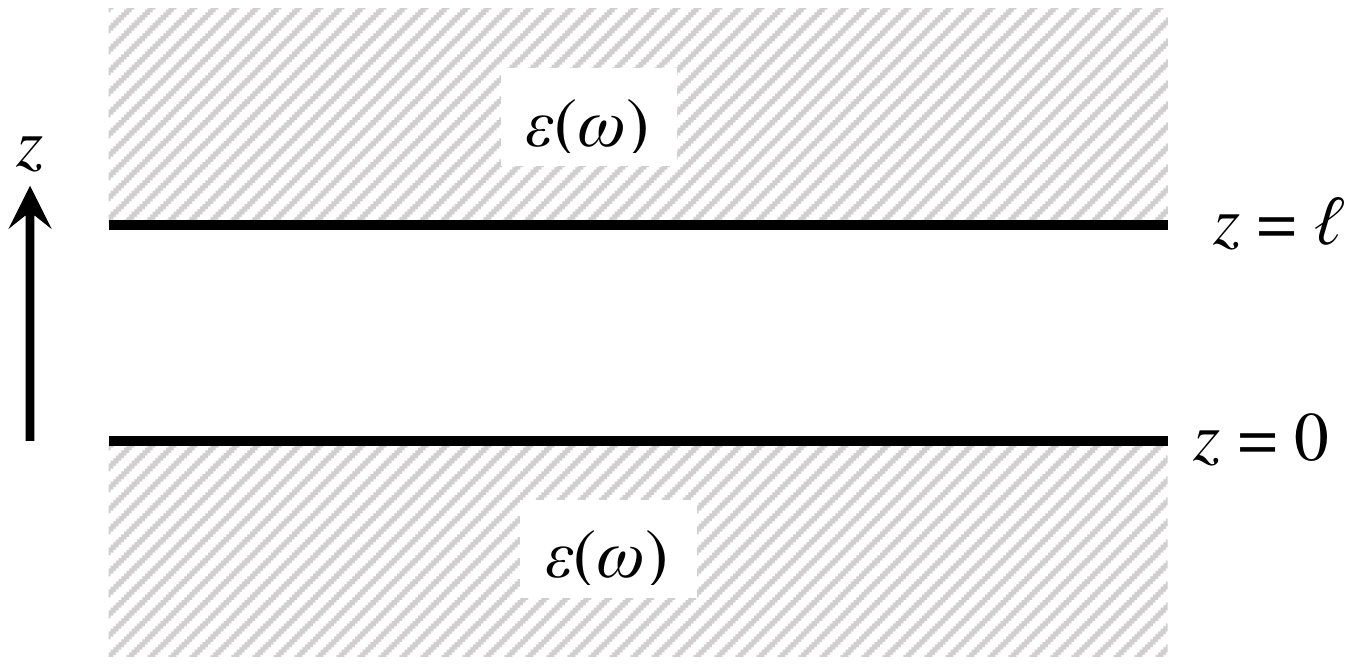}
\caption{Definition sketch}
\end{figure}

Since our aim is to focus on conceptual issues, we simplify the setup by assuming that the plates (half-spaces) are made of the same nonmagnetic material, with permittivity $\varepsilon(\omega)$. Furthermore, we  consider the nonretarded limit which formally amounts to setting the velocity of light to infinity. In this case it is sufficient to keep only the Poisson equation for the electric field ${\bf E}({\bf r},t)=
- \nabla\phi({\bf r},t)$, instead of the full set of Maxwell's equations. The necessary condition for neglecting retardation is that the width $l$ of the gap is smaller than  the electromagnetic wavelength at the relevant frequencies. The alternative way of stating this condition is that
 the time of light propagation over the distance of the gap width $l$ must be  smaller than the other relevant time, namely the period of oscillations $ \sim 1/\omega.$

 The Fourier component  $\phi_{\omega}({\bf r})$ of the potential  satisfies
\begin{equation}
-{\bf \nabla}\cdot [\varepsilon({\bf r},\omega){\bf \nabla}\phi_\omega({\bf r})] = 4\pi \rho_\omega({\bf r}), \label{1a}
\end{equation}
where $\rho_\omega({\bf r})$ is the Fourier component of the fluctuating charge density $\rho({\bf r},t)$. The function $\varepsilon({\bf r},\omega)$ is equal to 1 in the gap and is given by $\varepsilon(\omega)$ in the plates. The statistical properties of the fluctuating charges are determined by the correlation function $\langle \rho({\bf r},t)\rho({\bf r}',t')\rangle$, or by its  Fourier transform with respect to time, which defines the spectral density \cite{rytov89}
\begin{equation}
	\langle \rho({\bf r})\rho^*({\bf r}')\rangle_\omega = \frac{\hbar}{8\pi^2}\coth \left( \frac{\hbar \omega}{2T}\right) \nabla_r^2 \left[ \rm{Im}~\varepsilon({\rm r},\omega)\delta({\rm r-r'})\right], \label{2a}
\end{equation}
where the angular brackets  indicate a statistical average \cite{foot}.   The temperature $T$ is here given in energy units, as is quite common in the literature \cite{landau84}. Formally, it corresponds to setting the Boltzmann constant $k_B=1$. The subscript $\omega$ in (\ref{2a}) means that the averaged  quantity corresponds to a "spectral density", i.e. represents fluctuations resolved in frequency \cite{landau80a}.
Equation (\ref{2a}) constitutes the FDT for the fluctuating charges. The spontaneous charge fluctuations (the LHS in (\ref{2a})) are related to dissipation in the medium (imaginary part of the permittivity). Equations (\ref{1a}) and (\ref{2a}) enable one to compute the spectral density for the fluctuating potential, and similarly for the electric field components.

Define the Green function
\begin{equation}
-{\bf \nabla}\cdot [ \varepsilon ({\bf r},\omega)\nabla_r G_\omega({\bf r,r'})]= \delta( {\bf r-r'}). \label{3a}
\end{equation}
Then, using (\ref{2a}) and the identity
\begin{equation}
-\int d{\bf r} {\rm{Im}}\varepsilon({\bf r},\omega)\nabla_rG_\omega ({\bf r,r_1}) \cdot \nabla_rG_\omega^* ({\bf r,r_2})= {\rm{Im}}\,G_\omega ({\bf r_1,r_2}), \label{4a}
\end{equation}
one obtains
\begin{equation}
\langle \phi({\bf r})\phi^*({\bf r'})\rangle_\omega= -2\hbar \coth \left(\frac{\hbar \omega}{2T}\right){\rm Im}\, G_\omega ({\bf r,r'}).\label{5a}
\end{equation}
From (\ref{5a}) one can find the spectral density $\langle E_\alpha ({\bf r})E_\beta({\bf r'})\rangle_\omega$ for the electric field components, and then the average Maxwell stress tensor (\textbf{for the definition see e.g. \cite{landau84} \S5}).

For planar geometry it is easy to calculate the Green function explicitly, to obtain from (\ref{5a}) the $zz$- component of the stress tensor. The surface pressure becomes
\begin{align}
f= \int_{-\infty}^\infty \frac{d\omega}{2\pi}T_{zz}(l,\omega)= \int_{-\infty}^\infty \frac{d\omega}{2\pi}\frac{1}{8\pi}\left[\langle E_z^2(l)\rangle_\omega -2\langle E_x^2(l)\rangle_\omega \right] \nonumber \\
= \frac{\hbar}{4\pi^2}{\rm{Im}}\int_{-\infty}^\infty d\omega \coth \left( \frac{\hbar \omega}{2T}\right) \int_0^\infty \frac{q^2dq}{r^2(\omega)e^{2ql}-1}, \label{6a}
\end{align}
where
\[ r(\omega)= \frac{\varepsilon(\omega)+1}{\varepsilon(\omega)-1}. \]
The integral over $\omega$ is transformed to the complex $\omega$ plane with the help of of the "hairpin" contour (Fig. 2), to obtain
\begin{equation}
f= \frac{T}{8\pi l^3}{\sum_{n=0}^\infty}' \int_0^\infty \frac{x^2dx}{r^2(i\zeta_n)e^x-1}, \label{7a}
\end{equation}
where $\zeta_n= 2\pi nT/\hbar$. The prime means that the mode $n=0$ is taken with half weight.
Equation (\ref{7a}) gives the Casimir-Lifshitz force in the nonretarded limit (it does not contain the light velocity $c$).

\begin{figure}[t]
\includegraphics[width=4in]{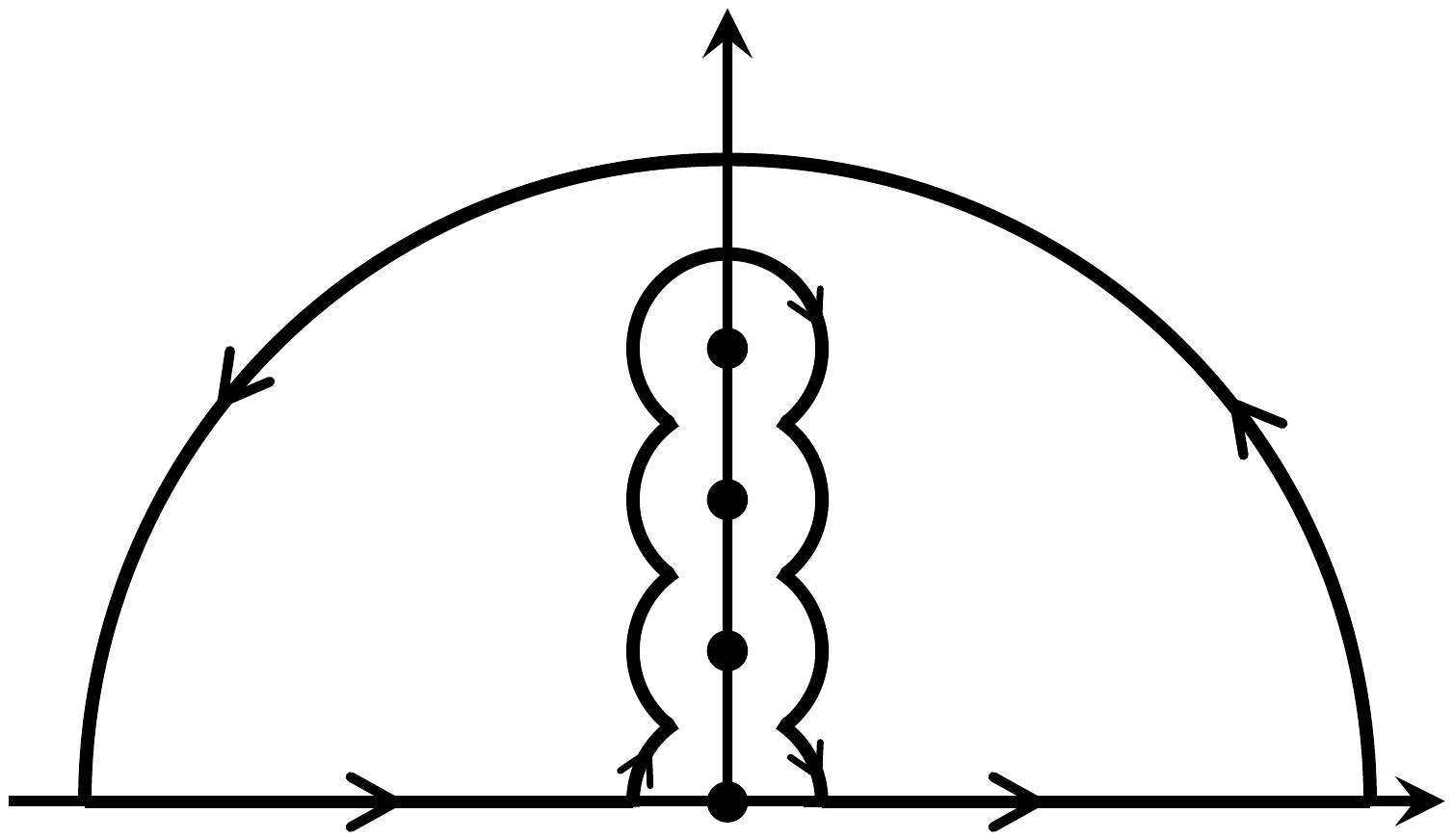}
\caption{Hairpin integration contour. The fat dots are the poles of the coth-functions. }
\end{figure}

A general property of $\varepsilon(\omega)$ is that \textbf{ lim$\,\varepsilon(\omega)|_{\omega \rightarrow\infty}=1$.}
 Let us introduce a characteristic frequency
$\omega_0$, beyond which $\varepsilon(\omega)$ rapidly approaches 1. [For a metal, $\omega_0 \sim\omega_p$ (plasma frequency). For a dielectric material, $\omega_0$ corresponds to a frequency region where strong absorption occurs, usually at optical frequencies; cf. \cite{landau80}.]

One can identify two temperature regions:

\noindent {\it (i) High temperature, $T  \gg\hbar \omega_0$}.

\noindent In this case  $\zeta_n \gg \omega_0$ for all $n$ (except $n=0$), so it is sufficient to  keep only the $n=0$ term,
\begin{equation}
f= \frac{T}{16\pi l^3}\int_0^\infty \frac{x^2dx}{r^2(0)e^x-1}. \label{8a}
\end{equation}
This is the classical limit (no $\hbar$).

\noindent {\it (ii) Low temperature, $T \ll \hbar \omega_0$}.

\noindent In this case many terms in (\ref{7a}) contribute to the sum, which can be replaced by an integral, leading to
\begin{equation}
f= \frac{\hbar}{16\pi^2l^3}\int_0^\infty d\zeta \int_0^\infty \frac{x^2dx}{r^2(i\zeta)e^x-1}. \label{9a}
\end{equation}
The last two equations correspond to Eqs.~(82.10) and (82.1) in \cite{landau80}.


\section{The QFT Approach}

In this approach the medium is considered as passive (i.e. no fluctuating currents in the medium) and the FDT theorem is not used. In fact, $\varepsilon({\bf r},\omega)$ is taken real and the aim is to find the modes of the electromagnetic field for the appropriate geometry. Each eigenfrequency is  assigned the corresponding thermal energy and the total free energy of the field (per unit area), $ F(l)$, as a function of the gap width $l$, is calculated. The Casimir-Lifshitz pressure is given by $f = -dF/dl$.

The QFT approach is an extension of the original calculation by Casimir \cite{casimir48} who considered the vacuum electromagnetic field between two ideal metallic surfaces. For two plates in Fig. 1, with real $\varepsilon(\omega)$, the QFT method was first employed by van Kampen et al. \cite{vankampen68} (see \cite{barash75}  for an early review).  In the nonretarded limit, considered in \cite{vankampen68}, one has to find the eigenmodes of the Poisson equation without sources
\begin{equation}
{\bf \nabla}\cdot [\varepsilon({\bf r}, \omega){\bf \nabla}\phi_\omega ({\bf r})=0. \label{11a}
\end{equation}
with the condition that the solution decays when $|z|\rightarrow \infty$.  This condition selects the surface modes which are the only relevant modes, due to their dependence on $l$.  A standard treatment \cite{vankampen68,barash75} leads to the following dispersion equation for the eigenfrequencies  of the surface modes:
\begin{equation}
g(\omega)\equiv 1-\frac{1}{r^2(\omega)}e^{-2ql}=0, \quad q= \sqrt{k_x^2+k_y^2}. \label{12a}
\end{equation}
The solutions $\omega_{\alpha}(k_x, k_y)$  are labeled by three indices:  the components $k_x, k_y$  of the transverse wave vector and the discrete number $\alpha$ that counts the solutions for fixed $(k_x, k_y)$. In the $ T\rightarrow 0$ limit, the free energy (per unit area) is
\begin{equation}
F(l)= \int \frac{dk_xdk_y}{(2\pi)^2}
\sum_\alpha \frac{1}{2} \hbar \omega_\alpha(k_x,k_y). \label{13a}
\end{equation}
The sum over $\alpha$, for fixed $(k_x, k_y)$, is performed with the help of the argument principle which states that for a meromorphic function $g(\omega)$, within some closed contour C in the complex-omega plane
\begin{equation}
\frac{1}{2\pi i}\oint d\omega \omega \frac{\partial}{\partial \omega}\ln g(\omega)= \sum_\alpha \omega_\alpha -\sum_\beta \omega_\beta, \label{14a}
\end{equation}
where $\omega_{\alpha}$ and $\omega_{\beta}$ are, respectively, the zeros and the poles of the function $g(\omega)$  within the integration contour.

\begin{figure}[th]
\includegraphics[width=4in]{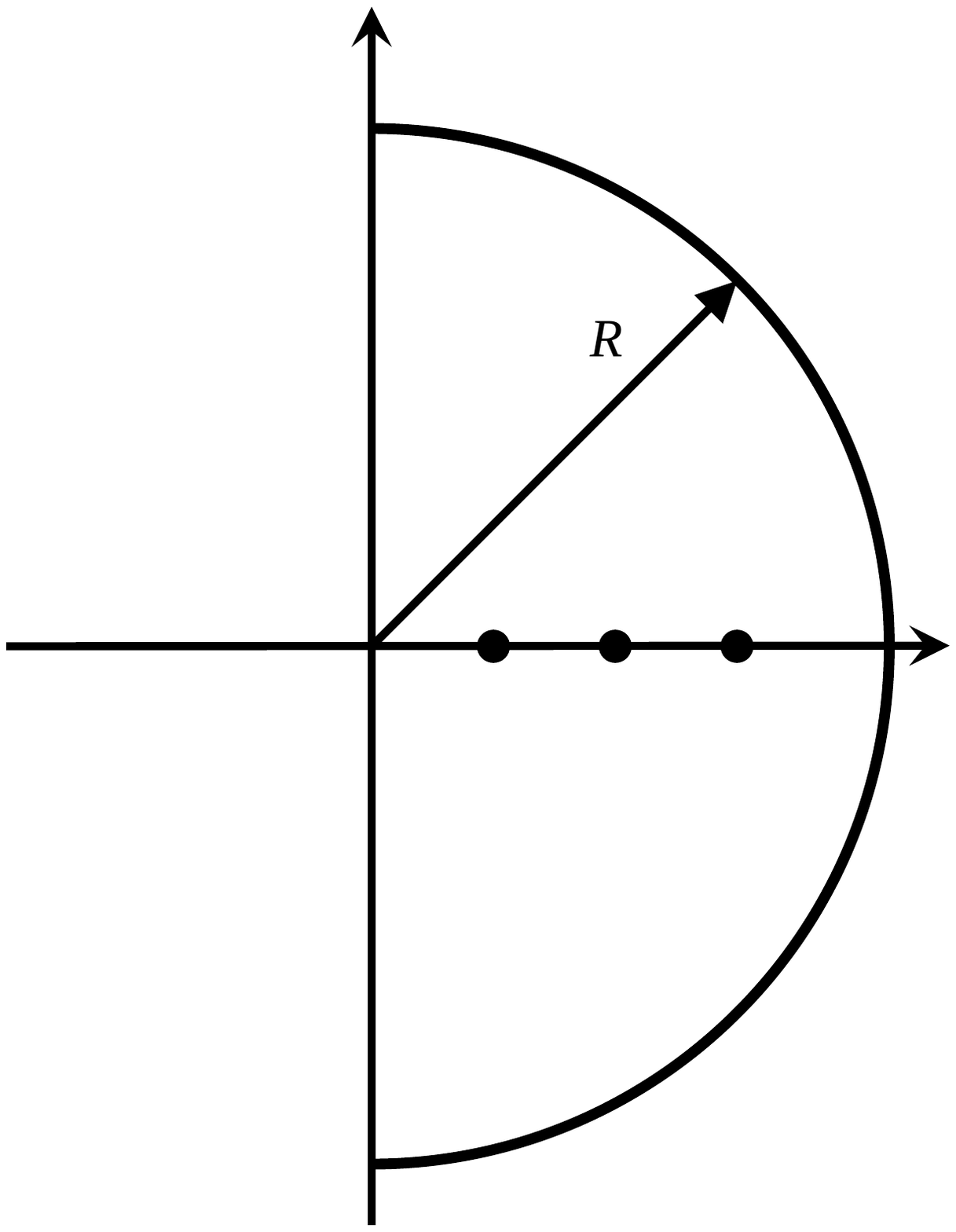}
\caption{The fat line gives the QFT integration contour.  The dots are the eigenfrequencies of the surface modes, $g(\omega)=0$.}
\end{figure}

 If we choose the contour shown in Fig 3, then Eq.~(\ref{14a})  (for $ R \rightarrow \infty$)  yields the desired sum of the eigenvalues, i.e. the roots of Eq.~(\ref{12a}).  (Note that the sum over the poles $\omega_{\beta}$ of $g(\omega)$  can be discarded because $\omega_{\beta}$  are the solutions of $r(\omega) = 0$ and, thus, cannot depend on $l$).  Since for $R \rightarrow \infty$   the integral along the semicircle does not contribute, we obtain
\begin{equation}
F(l)= -\frac{1}{2\pi i}\frac{\hbar}{2}\int \frac{dk_xdk_y}{(2\pi)^2}\int_{-i\infty} ^{i\infty} d\omega \omega \frac{\partial}{\partial \omega}\ln \left[ 1-\frac{1}{r^2(\omega)}e^{-2ql}\right], \label{15a}
\end{equation}
which, after integration by parts and relabeling the  variables, gives
\begin{equation}
F(l)= -\frac{\hbar}{8\pi^2}\int_0^\infty qdq \int_{-\infty}^\infty d\zeta
\ln \left[1-\frac{1}{r^2(i\zeta)}e^{-2ql} \right]. \label{16a}
\end{equation}
Finally, taking the derivative with respect to $l$, we obtain for the pressure
\begin{equation}
f= -\frac{\partial F(l)}{\partial l}= \frac{\hbar}{16\pi^2l^3}\int_0^\infty d\zeta \int_0^\infty \frac{x^2dx}{r^2(i\zeta)e^x-1}.  \label{17a}
\end{equation}
Formally, (\ref{17a}) looks exactly the same as (9), and we copy it only to stress that (\ref{17a}) was obtained under the initial assumption that $\epsilon(\omega)$, and thus the eigenvalues $\omega_{\alpha}$ must be real- otherwise the expression for the free energy, Eq.~(\ref{13a}) would make no sense.

 At first, there seems to be a logical contradiction: The FDT approach requires  $Im\varepsilon(\omega)>0$ (otherwise the force is identically zero) while the QFT approach assumes  $\varepsilon(\omega)=0.$ A closer look, however, reveals that also in the QFT approach one must introduce, at least implicitly, an infinitesimal  $Im\varepsilon(\omega)>0.$ In the above derivation this was done when the decaying (rather than growing) solutions were chosen for the surface modes. This is of course consistent with the general knowledge that a completely transparent medium, with no absorption at all,  is a fiction and in many situations one needs to add an infinitesimal (positive) Im$\varepsilon$  to make things well defined. This infinitesimal correction can be set to zero at the end, after the thermodynamic limit is taken (see e.g. \S77 of Ref.~\cite{landau80} where electromagnetic fluctuations in an infinite medium are considered). The same point is clearly stated in \cite{schwinger} where the authors, when using the QFT approach, select the Green's function which satisfies the boundary condition of an outgoing wave at infinity. This choice makes the integral over the real $\omega$-axis, which initially appears in the calculation of \cite{schwinger}, well defined and, moreover, it enables one to transform the integration to the imaginary semi-axis in the complex $\omega$-plane. The latter transformation is necessary when the theory is extended to finite $T$.

\section{Comparison between the Two Approaches}

As mentioned above, the
 two results  (\ref{9a}) and (\ref{17a}) look identical, although they have been derived under completely different, in fact opposite, conditions.
  In Eq.(\ref{9a}),  $\varepsilon(\omega)$ was required to have a finite imaginary part (as appropriate for realistic material bodies), while in the derivation of  Eq.(\ref{17a})  $\varepsilon(\omega)$ was taken as real (and only a closer inspection revealed that an infinitesimal  $Im\varepsilon(\omega)>0$ was needed to select the physical solutions for the eigenmodes). This second derivation (unlike the first one) is called in the review  \cite{barash75} "a prescription" rather than "a theory" and the prescription implies that, while integrating over the contour in Fig. 3, one should disregard possible singularities in the function $g(\omega)$ (or  $D(\omega)$ in the notations of  \cite{barash75}).


Although Eq.~(\ref{17a}) was derived under the very restrictive assumption of a transparent medium, it is tempting to use it also for absorbing media. Sometimes a plausible argument is being put forward \cite{ginzburg79}: Since integration in (\ref{17a}) is along the imaginary $\omega$-axis and since it is known that on this axis $\varepsilon(\omega)$ is real also for absorbing media, it is natural to extrapolate Eq.(\ref{17a}) to such media as well.
We know, of course, (by just looking at Eq (\ref{9a})) that such an extrapolation is indeed correct. However, much effort has been done in order to demonstrate this rigorously, staying solely within the QFT approach, with its mode counting procedure. Such rigorous considerations require an explicit introduction of a thermal bath, in equilibrium with the system consisting of the material bodies plus electromagnetic radiation (see \cite{behunin} and references therein).
It is not our purpose to argue about advantages and disadvantages of the two approaches or to dwell on the details of either of them (e.g. we don't even present the extension of the QFT approach to arbitrary temperatures, which is rather straightforward). Let us instead emphasize again the dichotomy between the two approaches:
The FDT approach makes it clear from the start that  fluctuating electromagnetic fields  originate in dissipative material bodies and that using a model with strictly real  $\varepsilon(\omega)$ (for real $\omega$) would make no sense, since there would not be any fluctuations. On the other hand, the starting point of the QFT approach is the electromagnetic field at thermal equilibrium, while the surrounding material bodies are "passive" (non-fluctuating) with strictly real  $\varepsilon(\omega)$. (Note that such non-dissipating bodies cannot establish thermal equilibrium with radiation!) Our main point here is that, although one can substitute a strictly real (for real $\omega$)  $\varepsilon(\omega)$ into Eqs.~ (\ref{9a}) and  (\ref{17a}) (or their generalization to an arbitrary temperature, Eq.~(\ref{7a})) and obtain meaningful results, one should keep in mind that the very derivation of those equations required presence of some finite (positive) Im$\varepsilon(\omega)$. Postulating a model with a strictly real  $\varepsilon(\omega)$ for all real $\omega$, can lead to contradictions and inconsistencies.

It is common knowledge that, although a strictly real  $\varepsilon(\omega)$ can be an acceptable approximation in some range of frequencies and material parameters, no realistic material can have a real  $\varepsilon(\omega)$ at all frequencies. Indeed, as clearly stated in the textbook \cite{landau84}, p.280 "...the imaginary part of $\varepsilon$ is positive for positive real omega  i.e. on the right-hand half of the real axis."  A more complete and general discussion, in terms of the susceptibility $\alpha(\omega$ (which differs from  $\varepsilon(\omega)$ just by a constant value 1) is given in \cite{landau80a}, p. 379, where it is emphasized "we reach the important conclusion that, for all positive values of the variable $\omega$, the function Im$\alpha(\omega)$  is positive and not zero". Thus, a material with strictly real  $\varepsilon(\omega)$, at all frequencies, is inadmissible. Indeed, such a material would violate the Kramers-Kronig relations, as well as some rigorous sum-rules (see Eq. (82.12) in \cite{landau84}:
\begin{equation}
\frac{m}{2\pi^2e^2}\int_0^\infty \omega {\rm Im}\varepsilon(\omega)d\omega =N, \label{18a}
\end{equation}
where $N$ is the electron concentration and ${e}$ and $ m$ are their charge and mass. It would also violate the rigorous relation between $\varepsilon(\omega)$ on the real and imaginary $\omega$ axes (see (82.15) in \cite{landau84};
\begin{equation}
\varepsilon(i\omega)-1 = \frac{2}{\pi}\int_0^\infty \frac{x{\rm Im}\varepsilon(x)}{x^2+\omega^2}dx. \label{19a}
\end{equation}
Furthermore, as explained in \cite{landau84},  an admissible $\varepsilon(\omega)$, in addition to being analytic in the upper half $\omega$ plane (causality), should have no zeroes in that half-plane (if $\varepsilon(\omega)$ is admissible, then so is $1/\varepsilon(\omega)$). As for the $\omega=0$ point, $\varepsilon(\omega)$ at this point can be either regular (dielectrics) or have a first order pole (metals) \cite{landau84}.

All these considerations notwithstanding, one can still find in the literature on Casimir-Lifshitz forces a model with strictly real $\varepsilon(\omega)$. This is the dissipation-less plasma model (DPM)\cite{mostepanenko21},  with
\begin{equation}
\varepsilon_p(\omega)= 1-\frac{\omega_p^2}{\omega^2}, \label{20a}
\end{equation}
where $\omega_p = (4\pi e^2 n_0/m)^{1/2}$ is the plasma frequency ($n_0$ is the electron concentration). This model is often considered on par with the Drude model
\begin{equation}
\varepsilon_D(\omega)= \varepsilon_L(\omega)-\frac{\omega_p^2}{\omega[ \omega + i\gamma(T)]}, \label{21a}
\end{equation}
which does allow for dissipation via the relaxation frequency $\gamma(T)$ which can depend on temperature $T$. The term $\varepsilon_L(\omega)$ accounts for the polarization of the lattice. In the low frequency limit, $\omega \ll\gamma$, one arrives to the often used expression
\begin{equation}
\varepsilon(\omega)= \varepsilon_L(0)+i\frac{4\pi \sigma(T)}{\omega}, \label{22a}
\end{equation}
where $\varepsilon_L(0)$ is the static limit of the lattice dielectric constant and $\sigma(T) = \omega_p^2/\gamma(T)$ is the dc conductivity of the mobile carriers. We will return to this expression in Sec. VI.

In addition to being dissipation-less, the plasma model in Eq.~(\ref{20a})  has the strange feature of exhibiting a second order pole at $\omega=0$, in contradiction with the above mentioned possibilities stated in  \cite{landau84} . This pole results in a puzzling discrepancy between the DPM and the Drude model, even if  in the latter $\gamma$ is taken arbitrary small \cite{mostepanenko21}. In this context the DPM has been already criticized in the literature (see e.g. \cite{brevik08}). Here we argue that the second order pole cannot exist in any realistic plasma (even as a meaningful approximation). Indeed, while Eq.~(\ref{20a})   can be a good approximation for a collision-less plasma at high frequencies     (and with Landau damping being neglected), it becomes completely meaningless near $\omega=0$. As extensively discussed, for instance, in the textbook \cite{lifshitzkinetics},   at low frequencies spatial dispersion becomes unavoidable so that Eq.~(\ref{20a})  fails completely and must be replaced by a (tensor) function $\varepsilon_{\alpha\beta}(\omega, {\bf k})$ depending on both frequency $\omega$ and wave vector ${\bf k})$. We will return to this point in the next section.

We close this general discussion by briefly mentioning the "generalized Kramers-Kronig relations" for the DPM, proposed in \cite{klimchitskaya07}. It is obvious that the DPM, as any model with strictly real $\varepsilon(\omega)$, violates the standard Kramers-Kronig relation. It suffices to consider one of the two relations, see e.g Eq.(82.6) in \cite{landau84}:
\begin{equation}
{\rm Re}\varepsilon(\omega)=1+\frac{1}{\pi}P \int_{-\infty}^\infty \frac{{\rm Im}\varepsilon(x)}{x-\omega}. \label{23a}
\end{equation}
Since (\ref{23a})  is clearly incompatible with (\ref{20a}),  the authors of \cite{klimchitskaya07}  propose a "generalized Kramers-Kronig relation " by simply subtracting a term $(\omega_p/\omega)^2$ from the RHS of (\ref{23a}). Such a subtraction yields a trivial identity for Re$\varepsilon(\omega)$, which lacks any physical content \cite{comment}. Furthermore, this sort of "generalization" could be proposed for any hypothetical substance with  arbitrary Re$\varepsilon(\omega)$ and Im$\varepsilon(\omega)=0$ (for all real positive $\omega)$. Such a substance obviously violates the Kremers-Kronig relation (\ref{23a}). One could try to "fix" this contradiction by adding a term [Re$\varepsilon(\omega) -1$] to the RHS of (\ref{23a}).  This would lead to a trivial identity for the specific Re$\varepsilon(\omega)$ of the hypothetical substance, which does not make any sense.  All such models with strictly real $\varepsilon(\omega)$ are inadmissible.

\section{ DPM, Drude and the Collision-less Plasma Model}

The DPM and the Drude model have been extensively studied in connection with the Casimir-Lifshitz forces. The most significant difference between the two is that for the Drude model
\begin{equation}
\lim \omega^2\varepsilon(\omega)|_{\omega \rightarrow 0} =0,\label{24a}
\end{equation}
while for the DPM this limit is finite, due to the second order pole at $\omega=0$. This discrepancy leads   to widely different results    for the Casimir-Lifshitz force in the two models. Below we examine closer the case when both $\omega$ and $\gamma$ become small.

When $\gamma=0$, i.e. all scattering mechanisms for the electrons (phonons, impurities or other electrons) are neglected, one arrives at the limit of what is called "collision-less plasma". However, generally (and for small $\omega$ in particular)  the dielectric function of such plasma does not at all follow Eq.~(\ref{20a}).   As explained in \cite{landau84,lifshitzkinetics},  in any conducting medium  the relation between the current density and the electric field is in general nonlocal, i.e. the current density at some point depends on the electric field in some vicinity of that point. The extent of this "vicinity" (the correlation radius $r_{cor}$) is determined by one of the two following mechanisms: (i) the scattering mean free path $l_{mfp}~ \sim \bar{v}/\gamma$, where $\bar{v}$ is the average electron velocity, or (ii) the length $l_{\omega} \sim \bar{v}/\omega$, which is the length over which an electron is displaced (in the absence of collisions) during one period of field oscillation. The correlation radius $r_{cor}$ is determined by the smallest of the two lengths. Since for the collision-less plasma $l_{mfp} \rightarrow \infty$, we have $r_{cor} \sim \bar{v}/\omega$. Only for $kr_{cor} \ll 1$, i.e. $\omega \gg k\bar{v}$, the local relation between current density and electric field is justified and the notion of $\varepsilon(\omega)$ becomes meaningful. In the opposite case, $\omega<k\bar{v}$, spatial dispersion becomes essential and the dielectric tensor $\varepsilon_{\alpha\beta}(\omega, {\bf k})$ must be used. Let us stress that even in a collision-less plasma $(\gamma=0)$, $ \varepsilon(\omega, {\bf k})$ contains an imaginary part, due to the Landau damping.

Microscopic treatment of spatial dispersion is based on a kinetic equation and the expressions for $\varepsilon_{\alpha\beta}(\omega, {\bf k})$ , for the collision-less plasma as well as for plasma with collisions, can be found in \cite{lifshitzkinetics}. Similar expressions, for somewhat different microscopic models, have been employed in \cite{svetovoy05,sernelius05} for calculating the Casimir-Lifshitz forces. Some phenomenological expressions for $ \varepsilon_{\alpha\beta}(\omega, {\bf k})$ have been written down in \cite{mostepanenko21},  Eq. (61), but those do not seem to have any microscopic justification.

We do not write down specific expressions for $ \varepsilon_{\alpha\beta}(\omega, {\bf k})$ for a collision-less plasma. Those can be found in \cite{lifshitzkinetics}, and depend on the temperature and electron concentration which in turn determine whether the plasma is degenerate (metal) or not (semiconductor). Let us only note that spatial dispersion removes not only the second order pole in Eq.~(\ref{20a})   but also the first order pole at $\omega=0$ which is characteristic of any conducting medium in the absence of spatial dispersion.

\section{ Drude Model with Spatial Dispersion}

The importance of spatial dispersion has been recognized already in the early work on fluctuational electrodynamics \cite{Levin67,rytov89}.  Let us illustrate this by a simple example. In the nonretarded limit (see Sec. II) the electrodynamic part of the problem amounts to the Poisson equation (\ref{1a}) but this time we must treat the dielectric function as a tensor  $\varepsilon_{\alpha\beta}(\omega, {\bf k})$. Assuming a homogeneous medium, transforming (\ref{1a})  to the $k$-space and introducing
\begin{equation}
\varepsilon(\omega, {\bf k})= \frac{1}{k^2}\sum_{\alpha,\beta}k_\alpha k_\beta \varepsilon_{\alpha\beta}(\omega, {\bf k}), \label{25}
\end{equation}
we obtain, for a homogeneous medium
\begin{equation}
k^2\varepsilon(\omega, {\bf k})\phi_{\omega}({\bf k})= 4\pi \rho_{\omega}({\bf k}), \label{26a}
\end{equation}
where $\rho_{\omega}({\bf k})$  designates the spontaneous fluctuation sources in the $(\omega, {\bf k})$ representation. The dielectric function  $\varepsilon_(\omega, {\bf k})$  is called the longitudinal dielectric function and  is usually designated by a subscript $l$. Since, however, this is the only dielectric function relevant in our treatment, we omit this subscript. It follows from Eq.~(\ref{2a})  (generalized to the case of spatial dispersion)  that the spectral density
\begin{equation}
\langle \rho({\bf k})\rho^*({\bf k'})\rangle_\omega = \frac{\hbar k^2}{8\pi^2}(2\pi)^3\varepsilon(\omega, {\bf k})\coth \left(\frac{\hbar \omega}{2T}\right)\delta (\bf{k-k'}). \label{27a}
\end{equation}
Equations (\ref{26a}) and (\ref{27a})  enable one to study electric field fluctuations in a medium with any specified $\varepsilon_{\alpha\beta}(\omega, {\bf k})$ \cite{rytov89,Levin67,shapiro10}.

Let us now present the calculation of $\epsilon(\omega, {\bf k})$ for a simple model based on a hydrodynamic equation for a plasma (see e.g. \cite{shapiro10,dalvit08,davies}). The equation of motion is
\begin{equation}
m\frac{\partial{\bf v}}{\partial t}= e{\bf E}   -m\gamma {\bf v}-\frac{1}{n_0} {{\bf{\nabla}} p}, \label{28a}
\end{equation}
where ${\bf v}({\bf r}, t)$ is the plasma velocity at point $\bf r $ at time $t, n_0 $ is the equilibrium concentration of carriers,  $n({\bf r}, t) $ is the deviation from equilibrium, and $p = [n_0 +n({\bf r}, t)]T $ is the thermal pressure.  Equation (\ref{28a}) is based on the classical Boltzmann equation with Maxwell statistics for the carriers, which  provides a good description of a low density plasma in semiconductors. It is already linearized with respect to $\bf v $ and $n$, and it should be supplemented by a (linearized) continuity equation
\begin{equation}
\frac{\partial n}{\partial t} + n_0{\bf \nabla}\cdot {\bf v}=0. \label{29a}
\end{equation}
Fourier transforming    (\ref{28a}) and (\ref{29a}),  both in time and space, one can relate the current density ${\bf j}(\omega, {\bf k}) = en_0 {\bf v}(\omega, {\bf k}) $ to the electric field ${\bf E}(\omega, {\bf k}),$ thus obtaining the conductivity tensor and, ultimately the dielectric function
\begin{equation}
\varepsilon(\omega, {\bf k})= \varepsilon_L(\omega)- \frac{\omega_p^2}{\omega(\omega+i\gamma)-k^2R_D^2\omega_p^2}, \label{30a}
\end{equation}
where at the last stage the lattice contribution $\varepsilon_L(\omega) $ has been added. $ R_D = (T/m\omega_p^2)^{1/2}$  in (\ref{30a})   is the Debye screening radius. The importance of screening in the Lifshitz theory for conductors with low electron concentration has been emphasized in \cite{dalvit08,pitaevskii08}.  If spatial dispersion in (\ref{30a}) is neglected, the standard Drude model, Eq.~(\ref{21a}) is recovered. Note that spatial dispersion completely obliterates the pole at $\omega=0$ in the Drude model  (as well as, of course,  the second order pole in the DPM, Eq.~(\ref{20a})). In the low frequency limit (\ref{30a}) reduces to
\begin{equation}
\varepsilon(\omega, {\bf k})= \varepsilon_L(0) + \frac{i4\pi \sigma(T)}{\omega+ i4\pi \sigma(T)k^2R_D^2}, \label{31a}
\end{equation}
which should be compared to Eq.~(\ref{22a})  in the absence of spatial dispersion. Note that (\ref{22a})  exhibits a peculiar behavior in the limit of small frequency and low temperature. Since in a semiconductor (or, in fact, any dielectric material) $\sigma(T)$ rapidly approaches zero when $ T \rightarrow 0$,  we have at $T=0$,  $\varepsilon(\omega, k=0)=\varepsilon_L(0)$ for any finite $\omega$.  On the other hand, for $T $ different from zero and $\omega \rightarrow 0$, Eq.~(\ref{22a})  yields $\varepsilon = \infty$. Thus, in the absence of spatial dispersion, the two limits , $T \rightarrow 0$ and $\omega \rightarrow 0$, do not commute. This discontinuity leads to various problems in the theory of Casimir-Lifshitz forces and, in particular, to the so called "Casimir conundrum" \cite{mostepanenko21}, which amounts to violation of the Nernst heat theorem. It has been already mentioned in the literature \cite{pitaevskii08,milton12} that this unphysical "conundrum" is due to the inappropriate neglect of spatial dispersion. Indeed, spatial dispersion is known to smear out various singularities. For instance, the singularity that exists in the correlation function for the fluctuating thermal electric field in an infinite medium \cite{landau80}, gets regularized when spatial dispersion is accounted for \cite{Levin67}. Similarly, the effect of the above mentioned discontinuity, leading to the "conundrum", will disappear when the dielectric function (\ref{31a})  is used, with the subsequent integration over the transverse part of $ \bf k$. Indeed, the infinite jump of $\varepsilon$, just mentioned above in connection with Eq.~(\ref{22a}), disappears due to the second term in the denominator of (\ref{31a}). Furthermore,
 if one takes the formal limit $R_D \rightarrow \infty$ when $T $ (and thus $n_0, \omega_p$ and $\sigma)$ approaches zero, one finds that the $\omega \rightarrow 0$ limit does commute with the   $ T \rightarrow 0$ limit so there is no reason at all to suspect any "conundrum".
These qualitative arguments are, of course, not a substitute for a rigorous calculation of the force, using the expression (\ref{31a}) for the dielectric function.

\section{Final remarks}

Our main purpose was to juxtapose two approaches in the theory of the Casimir-Lifshitz forces, using the simple example of the force in the non-retarded limit. In the FDT approach, with its emphasis on the fluctuating currents in the media as sources of the fields and forces, it is immediately obvious that dissipation in the media is indispensable. On the other hand, the QFT approach is based on finding the electromagnetic modes, while the surrounding medium is considered as passive and dissipation-less. For instance, in the monograph \cite{bordag09}, written apparently by the devotees of the QFT approach, in  their derivation of the Lifshitz formula  the authors state (p.287): "In the above derivation, the small imaginary parts of the photon eigenfrequencies were neglected". However, if indeed one could derive the Lifshitz formula for a medium with strictly real $\varepsilon(\omega)$ (i.e. no dissipation at all), it would lead to an immediate conflict with the FDT approach which yields an identically zero force in the absence of any dissipation. This apparent discrepancy is resolved if one realizes that also in the QFT approach  an infinitesimal (positive)  $Im\varepsilon(\omega)$ must be introduced, perhaps implicitly.

The necessity of introducing an  $Im\varepsilon(\omega)>0$ has not been enough emphasized in the QFT approach, which prompted some authors to advocate models without any, even infinitesimal, $Im\varepsilon(\omega)$. Such models violate some requirements that any realistic material must fulfill, like the Kramers-Kronig relations or some exact sum-rules. In particular, we criticize the dissipation-less plasma model (DPM) with its nonphysical and completely artificial second order pole in $\varepsilon(\omega)$ at $\omega=0$. It appears, though, that in recent years the popularity of the DPM is diminishing and that even its prime promoter, the author of the review \cite{mostepanenko21}, does not seem to strongly insist that this model is realistic. But, nevertheless, a significant portion of that review is devoted to DPM and one finds claims that  the model is "quite reasonable from the theoretical point of view" or that it has an advantage over the Drude model for metals  (\ref{21a}) because the latter violates the Nernst heat theorem. More precisely, this violation occurs only if it is assumed that $\gamma(T)$ in (\ref{21a}) approaches zero faster than linearly with $T$, and it is known as "Casimir puzzle" \cite{mostepanenko21} (to be distinguished from the "Casimir conundrum" mentioned in the previous section). In our opinion, neither the "success" of the DPM nor the "failure" of the Drude model have much meaning. The point is that neither the DPM nor the Drude model (with  $\gamma(T)$  rapidly approaching zero when $ T \rightarrow 0$ )  are applicable at low $T$ and small $\omega$, when the spatial dispersion must be taken into account. Moreover, in a real metal the zero temperature limit of $\gamma(T)$ is not zero but is some constant $\gamma_i$ which accounts for the residual scattering rate on static impurities. It was pointed out in \cite{hoye03,brevik06,hoye07} that for $\gamma_i$ different from zero the Drude model is entirely consistent with thermodynamics. Thus the "Casimir puzzle" can be resolved either by taking into account spatial dispersion (if it is  assumed that  $\gamma(T)$ rapidly approaches zero at low temperatures)
 or by simply recalling that in a real metal there is always some concentration of static impurities. Generally, both mechanisms can contribute simultaneously.

  In conclusion, one cannot have a reliable theory of the Casimir-Lifshitz forces for real materials unless one solves the problem for a realistic model, spatial dispersion included. A reliable theory must use reliable models! We cannot agree with a statement like "The Lifshitz theory is experimentally consistent only if one ignores the real physical phenomenon- small but quite measurable electric conductivity"\cite{mostepanenko21}. One cannot make a theory "consistent" by neglecting relevant physical phenomena. We conclude with an amusing historical note:

In the 1970's, during a visit that Hendrik B. G. Casimir made to the Institute of Theoretical Physics at the Norwegian Institute of Technology in Trondheim, one of the authors (I.B.), then an assistant at the institute,  attended a lecture that our guest  gave on a topic quite different from what has later been known as the Casimir effect. At that time  the effect was actually not very well known, but in some way I had gotten to know about the effect.  In the discussion session after the lecture, I asked: "Is the Casimir effect due to the quantum mechanical field fluctuations, or is it due to the van der Waals forces between the molecules in the media?"  Casimir's  answer began as follows: "I have not made up my mind".
This answer may be the most precise answer that can be given even today. The dichotomy  of the effect, as we have tried  to elucidate above, is one of its most characteristic properties.

\bigskip

ACKNOWLEDGMENTS:

B. S. acknowledges useful correspondence with V. M. Mostepanenko, as well as enlightening discussions with Joshua Feinberg. We also thank Reidar Kristoffersen for help with the figures.


\begin{thebibliography}{99}
\bibitem{foot} Strictly speaking, in \cite{rytov89} one finds the spectral density for currents, rather than charges, but the two are related by the continuity equation.
\bibitem{comment} In \cite{klimchitskaya07} another term, $A(\omega)$, is added to the DPM  of Eq.~(\ref{20a}). That additional term describes "core electrons", which have nothing to do with the proposed "generalization of the Kramers-Kronig relation". That generalization  was needed to "fix" the obvious contradiction between the DPM and the standard Kramers-Kronig relation.
\end{thebibliography}
\end{document}